\UseRawInputEncoding
\documentclass[
 reprint,
superscriptaddress,
 amsmath,amssymb,
 aps,
prl,
]{revtex4-1}

\usepackage{graphicx}
\usepackage{dcolumn}
\usepackage{bm}
\usepackage{hyperref}

\def\v#1{{\bf#1}}
\def\be{\begin{equation}}
\def\ee{\end{equation}}
\def\bea{\begin{eqnarray}}
\def\eea{\end{eqnarray}}

\def\ncal{\mbox{$\cal N\,$}}
\def\<{\langle}
\def\>{\rangle}

\begin{document}


\title{Does an accelerated mirror suffer hindrance from vacuum?}

\author{E. Sadurn\'i}
 \email{sadurni@ifuap.buap.mx}
\affiliation{Instituto de F\'isica, Benem\'erita Universidad Aut\'onoma de Puebla,
Apartado Postal J-48, 72570 Puebla, M\'exico}

\author{M. A. Est\'evez}
\affiliation{Instituto de F\'isica, Benem\'erita Universidad Aut\'onoma de Puebla,
Apartado Postal J-48, 72570 Puebla, M\'exico}

\author{J. L. D\'iaz-Cruz}
\affiliation{Facultad de Ciencias F\'isico Matem\'aticas, Benem\'erita Universidad Aut\'onoma de Puebla, 72570 Puebla, M\'exico}


\begin{abstract}
We study the accelerated motion of mirrors and the photonic Unruh effect. The solutions of Maxwell's equations with appropriate boundary conditions in Rindler coordinates are found. The canonical quantization of the field is carried out properly. Important consequences arise from the presence of mirrors or polarizers. It is shown that a single reflective surface produces frequency quantization due to Dirichlet conditions applied to the $-1/(\alpha x)^2$ anomalous potential for non-inertial observers. The number of photons per accelerated mode is estimated via Hankel functions. \end{abstract}



\maketitle

Accelerated mirrors in a flat background possess a notion of quantum vacuum that differs significantly from that of inertial observers. Such is the lesson that we have learnt from the Unruh effect, originally studied for scalar particles \cite{unruh1976, crispino2008}. By studying the mechanism of particle production --Bogoliubov transformation-- we may conclude that the outcome should be similar for other types of quantized fields, either bosonic or fermionic, including Maxwell's theory \cite{gill2010, narozhny2001, tulipant2021, schleich2019}. Therefore, an accelerated mirror or polarizer interacting with the inertial vacuum state must feel an effective radiation field, and it should modify its motion as a reaction to radiation pressure. In other words, an accelerated object collides with real photons, damping its motion. Similar statements would apply when the mirror moves along the geodesics of a background curved space, e.g. a blackhole. 

The Unruh effect has spurred many experimental \cite{jiaz2019, uliana2019, lynch2020} and theoretical \cite{carballo2019, cisco2020, ivette2011, vanzella2017} investigations on field quantization in curved spaces, as well as some polemical discussions on plausible detection schemes \cite{ford2006}. For gravitational effects on interference patterns, see \cite{condado2018}. Although these discussions delve deep into the very nature of particle detectors in motion, it should be clear that quantum-mechanical averages also admit the interpretation of a fixed --previously prepared-- state and a transformed observable that belongs to the non-inertial observer. In this case, the observable is the particle number, and it is transformed together with the corresponding detector to the non-inertial frame. Therefore, if an accelerated observer starts abruptly its motion from rest and without visible radiation, an effective state preparation at inertial vacuum will take place. Moreover, if this happens at the beginning of our {\it gedankenexperiment,\ }the accelerated trajectory will force the detectors to operate on the previous vacuum state and not on the new one. The overall perception is that the transformed observable $U^{\dagger}N U$ will produce non-vanishing averages $\<0|U^{\dagger} N U|0\> \neq 0$. In the present work we are interested in studying the Unruh effect for photons interacting with mirrors, or polarizers, and solving the corresponding mathematical intricacies of such scenario, which to our knowledge has not been done before. In particular, we identify the corresponding boundary conditions and present a consistent solution (exact and approximated), which turns out to be non-trivial. We also discuss the physical interpretation of our findings. In our exploration, we shall find the existence of continuous and discrete frequency spectra of states localized by $-1/x^2$ potentials, also known as {\it quantum anomalous\ }interactions with fall-to-the-origin behavior \cite{tkachuk}, shape invariance under scale transformations \cite{serg} and Efimov states \cite{efimov}. These potentials also appear naturally in black hole horizons \cite{govin2000}. We shall verify this result in the corresponding wave equation after the use of Rindler coordinates and the Rindler metric, but with a slight variation in the case of photons as opposed to scalars. In the former case, the radiation gauge can be applied to the vector potential with the aim of eliminating polarization dependent interactions. Boundary conditions describing the position of one or two mirrors facing each other will also play an important role, as their presence will modify the nature of the frequency spectrum, i.e. from continuous to point-like. In this sense, accelerated Fabry-P\'erot interferometers can be regarded as photonic traps or bottles with a natural infrarred cutoff, providing thus the ideal conditions for a detector. In addition, photons do not have a natural length scale, such as $\lambda_{\rm Compton} = \hbar/mc$. As a result, non-inertial observers need not surpass large critical accelerations $\alpha_c$ in order to produce massless particles. The critical acceleration scale is the overwhelming quantity $mc^3/\hbar$ (for the Higgs boson, the acquired mass $m$ yields $\alpha_c \sim 10^{34}$m$/$s$^2$ and a more conservative example such as the pion yields $\alpha_c \sim 10^{31}$m$/$s$^2$). 

\begin{figure}
\includegraphics[width=8cm]{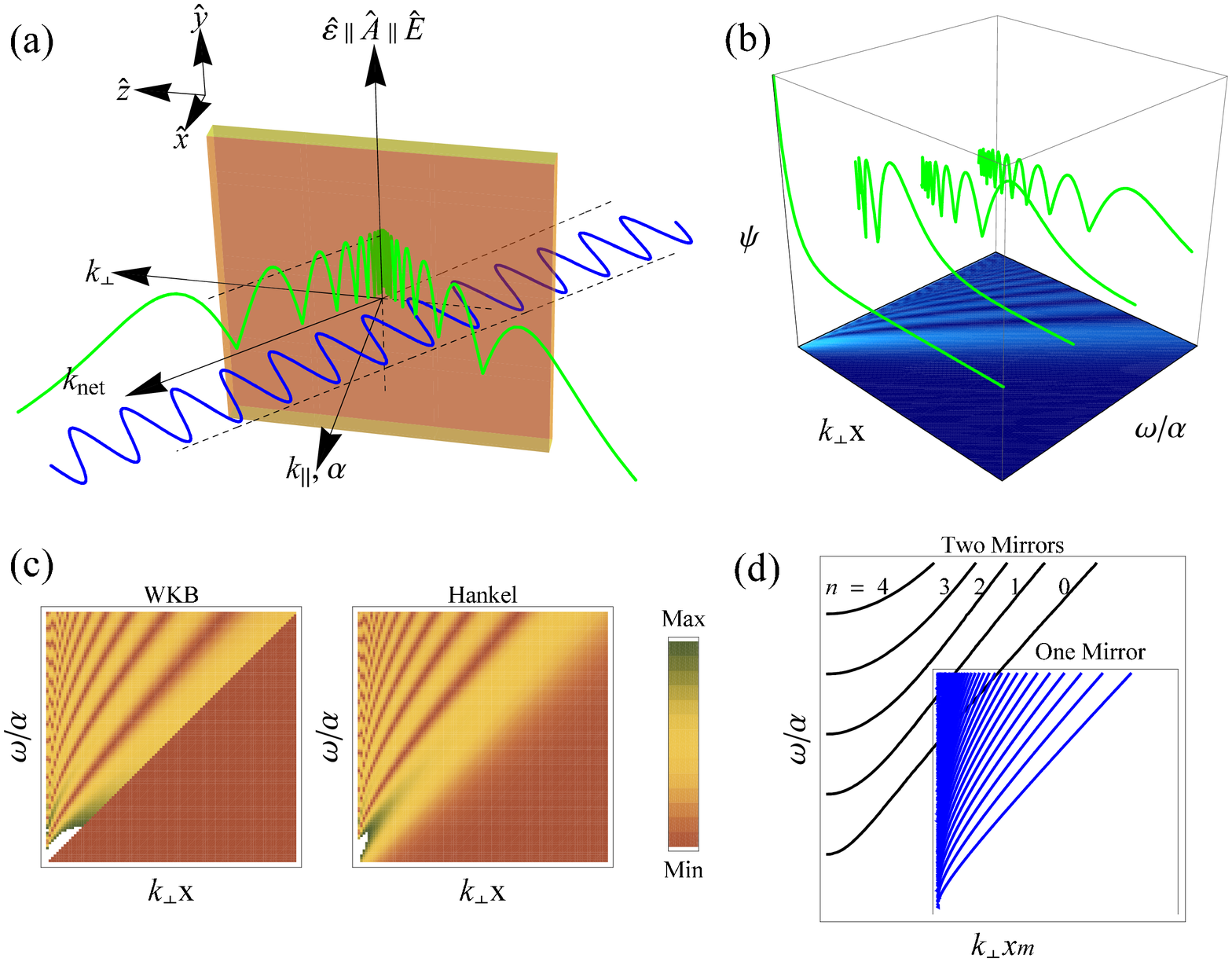}
\caption{\label{fig1} (a) Frontal view of an accelerated mirror with polarization axis $\hat z$ and perfect reflectivity along $\hat x$. The green signal corresponds to the chirped wave given by Hankel functions; the blue signal passes by. (b) and (c) Hankel functions for several values of $\omega/\alpha$, green curves. A comparison between densities $|H^{(1)}_{iy}(ix)|^2$ and the approximation (\ref{e12}) is included. (d) Frequency quantization (\ref{e5}) and (\ref{e6}) for one and two mirrors.}
\end{figure}
  
We start our treatment with the following notation (units $\hbar=c=1$). The Rindler coordinates of a right wedge involving $t$ and $x$ only, are given by 
\bea
T= x \sinh \alpha t, \quad X = x \cosh \alpha t,
\label{10}
\eea
and the corresponding metric is
\bea
g_{00} = \alpha^2 x^2, \quad g_{ij} = - \delta_{ij}, \quad g_{0i}=0.
\label{11}
\eea
The world line of an accelerated object in $(T,X)$ spacetime is described by $x=$ constant and $0<t<\infty$. In this respect, we shall always remember that in this system, the low acceleration limit is reached by letting $\alpha \rightarrow 0$ as well as $x \rightarrow 1/\alpha$, which corresponds to the asymptotically flat region $\alpha x=1$. Some authors prefer the redefinition $x \mapsto x + 1/\alpha$ in order to recover the flat limit seamlessly, but this is merely a translation in $x$. The Klein-Gordon equation has at least two equivalent forms (here, the scalar curvature is $R=0$) 

\bea
\left[ (|g|^{-1/2}) \partial_{\mu}(|g|^{1/2} g^{\mu \nu} \partial_{\nu}) + m^2 \right] \phi = 0, \nonumber \\
\left[\square + M^2(x) \right] \psi = 0,
\label{KG}
\eea
where $\square \equiv \partial_{\mu}g^{\mu \nu}\partial_{\nu}$ (not to be confused with $\nabla_{\mu}\nabla^{\mu}$) and $\psi = |g|^{1/4} \phi$. The second form can be useful, as it allows to understand the acceleration effect as a redefinition of the mass $M^2 = m^2 - 1/(2x)^2$, while the wave $\psi$ incorportates automatically the square root of the Jacobian. The wave equation coming from Maxwell's theory suffers a slightly different modification due to the new curvilinear coordinates, as it deals with a vector field instead of a scalar. We have an extra term coupled to polarization in the following equation

\bea
\frac{1}{\sqrt{g}} \partial_{\mu} \left(\sqrt{g} \partial^{\mu} A_{\nu} \right) - \Gamma^{\sigma}_{\mu \nu} \partial^{\mu} A_{\sigma} = 0,
\label{e1}
\eea
where we have made the choice of a covariant Lorentz gauge, i.e. $\nabla_{\mu}A^{\mu}=0 = (1/\sqrt{g}) \partial_{\mu}(\sqrt{g}A^{\mu})$. Now we solve (\ref{e1}) by separation of variables; the Lorentz gauge helps to eliminate the $\Gamma$- dependent polarization terms. In Rindler spacetime $\Gamma^{1}_{00}=\alpha x$, $\Gamma^{0}_{10}=\Gamma^{0}_{01}=1/(\alpha x)$ and all other Christoffel symbols vanish. Therefore

\bea
-\Gamma^{\sigma}_{\mu \nu} \partial^{\mu} A_{\sigma} = \left(-\alpha x \partial^{0}A_1 -\frac{\partial^{1}A_0}{\alpha x},-\frac{\partial^{0}A_0}{\alpha x},0,0\right).
\label{e2}
\eea
Meanwhile, separable solutions of the form $A_{\mu}(x_{\nu})= 
e^{-i\omega t} F_{\mu}(x) e^{i \v k_{\perp} \cdot \v x}$ can
be proposed. In fig. \ref{fig1}(a), we introduce the notation $\v k_{\rm net}=\v k_{\perp} + \v k_{\parallel}$ as a total wave vector, where $\v k_{\perp} \cdot \v x = k_y y+ k_z z$ is perpendicular to the accelerated motion and $\v k_{\parallel} = \hat x k_{\parallel}$ is defined along the mirror's trajectory, with magnitude $k_{\parallel} \equiv \sqrt{\omega^2 - k_{\perp}^2}$. For linear polarizers along $\v k_{\perp}$, a field polarization vector $\hat \varepsilon \parallel \hat y$ suffers reflection, which is the situation of interest. 

Upon substitution of $A_{\nu}$ into $\partial_{\mu}(\sqrt{g}g^{\mu \nu} A_{\nu})=0$ we find the usual transverse condition $A_0=A_1=0$ (no longitudinal photons for rectilinear accelerated motion) and the field is reduced to

\bea
A_2 = A_2(0) e^{-i\omega t}e^{i \v k_{\perp} \cdot \v x} F(x), \nonumber \\
A_3 = A_3(0) e^{-i\omega t}e^{i \v k_{\perp} \cdot \v x} F(x), \nonumber \\
\v k_{\perp} \cdot \v A = k_y A_2(0) + k_z A_3(0) = 0.
\label{e3}
\eea
As expected, the polarization of $\v A$ (and $\v E$) in the $(y,z)$ plane is also orthogonal to oblique wave vectors for this set of solutions, as shown in the green chirped pulses in fig. \ref{fig1}(a). It is important to note that there is another set of solutions for which $A_1 \neq 0$, leading to the gauge restriction $\partial_1 (x A_1)- i x \v k \cdot \v A =0$ and a non-vanishing $\Gamma$ term in (\ref{e1}). These solutions correspond to polarizations in the $x$ axis. If a beam splitter with perfect reflectivity for $y$-polarized waves is accelerated, only (\ref{e3}) will have the chance to interact with it. The other type of photons will pass by undetected, and they shall be ignored in what follows for simplicity.

Finally, we can reduce Maxwell's wave equation for (\ref{e3}) to the massless K.-G. theory in curved space for each component of $A$, i.e. $\left[ \square -1/(4x^2)\right] |g|^{1/4} A_{\mu} =0$. With this, the undetermined function $F(x)$, defined through $\psi = |g|^{1/4}F$, is shown to satisfy

\bea
\left[- \frac{d^2}{dx^2} - \frac{(\omega/\alpha)^2+1/4}{x^2} \right] \psi = - k_{\perp}^2 \psi.
\label{e4}
\eea
Indeed, the quantum anomalous potential along the direction of motion appears in (\ref{e4}). It is identical to a radial problem $-1/r^2$ for S waves; many important mathematical results have been obtained for this kind of eigenvalue problem. They are indispensable, as the completeness of the accelerated modes is required in the computation of relativistic inner products, which in turn play the role of coefficients in the Bogoliubov transformation. One of such properties is the scale invariance of the eigenvalue equation under $x \mapsto \lambda x, k_{\perp} \mapsto k_{\perp}/\lambda$, for any $\lambda \neq 0$, leading to a continuous spectrum in the absence of additional boundary conditions.  

A more delicate analysis begins here: It is well known that $F$ can be put in terms of Bessel functions with the help of the definitions $\nu \equiv i \omega/\alpha$, $\zeta \equiv i k_{\perp} x$ for the order $\nu$ and the argument $\zeta$ (both are imaginary). The Hankel functions are better suited for this study, as we can identify immediately that $H_{\nu}^{(1)}(i k_{\perp}x)$ satisfies the condition $|F|\rightarrow 0$ as $x \rightarrow \infty$, which is the only possibility for appropriate quantum-mechanical interpretations. This boundary condition at infinity {\it does not\ }quantize the frequency, as we can see from the scaling properties above. The effect of this restriction reduces to the elimination of $H^{(2)}_{\nu}(i k_{\perp}x)$, for it grows exponentially as $x\rightarrow \infty$.

Now we arrive at the crucial point; if a mirror accelerates {\it in vacuo,\ }the world line $x=x_{\rm m}=$ constant imposes a boundary condition on the photon's wave function precisely at this location. A mirror may interact with the produced radiation field in many forms, but if we think of this detector as a scatterer, there will be a resonance of the photon field when the phase shift of $\pi$ is attained. The reflective properties of a polarizer are not too different, if a specific polarization is chosen. The wave function at rest is $\sin\left[q (x-x_{\rm m})\right]$ for $x\geq x_m$ and $0$ otherwise. For the accelerated observer, the Dirichlet boundary condition still holds, but the wave function is now $\psi = \ncal H^{(1)}_{\nu}(i k_{\perp} x)$, so we must have the following {\it exact\ } quantization condition:  

\bea
H^{(1)}_{i \omega/\alpha}(i k_{\perp} x_{\rm m}) = 0, \, \, k_{\perp} x_{\rm m} = r_{n}(\nu), \, \, n=0,1,2,...
\label{e5}
\eea
where $r_n(\nu)$ are the (real) roots of the Hankel function of imaginary variable and index. A density plot in fig. \ref{fig1}(b) and (c) reveals the shape of these functions.

In this kind of anomalous potential, it is even more striking that a second boundary condition does not lead to an overspecified system. A photonic 'bottle' can be built with two mirrors facing each other at positions $x_1, x_2$. Only certain modes can be established inside of this device, which corresponds to a Fabry-P\'erot interferometer in motion (initially empty). The reason for frequency quantization is the allowed presence of $H^{(2)}$, as the full wavefunction vanishes identically outside of the bottle and becomes square integrable in the domain $x_1<x< x_2$. The eigenmodes and exact quantization condition are, respectively,

\bea
\psi &=& \ncal \sqrt{x} \left[ H_{i\omega/\alpha}^{(1)}(ik_{\perp}x_1) H_{i\omega/\alpha}^{(2)}(ik_{\perp}x)  \right. \nonumber \\ &-& \left.
H_{i\omega/\alpha}^{(2)}(ik_{\perp}x_1) H_{i\omega/\alpha}^{(1)}(ik_{\perp}x) \right], \nonumber \\
0&=&H_{i\omega/\alpha}^{(1)}(ik_{\perp}x_1) H_{i\omega/\alpha}^{(2)}(ik_{\perp}x_2) \nonumber \\ &-&
H_{i\omega/\alpha}^{(2)}(ik_{\perp}x_1) H_{i\omega/\alpha}^{(1)}(ik_{\perp}x_2).
\label{e6}
\eea  
Now we proceed to quantize the e.m. field that interacts with the mirror. Proper care must be taken in the discussion of polarization and Lorentz invariance (Gupta-Bleuler quantization). Firstly, from the superposition of solutions parallel and orthogonal to the mirror, i.e. $\v A_{\rm total} = \v A_{\parallel} + \v A_{\perp}$, we make sure that the field bouncing off the mirror (or inside the interferometer) is $\v A_{\parallel} \equiv \v A$. Secondly, the electric field is recovered here with $\v E = -\partial_t  \v A$ after $A_0=0$ has been chosen, so the field 'coordinate' and the field 'momentum' share the same polarization. Thirdly, the travelling mirror breaks both spatial isotropy and global Lorentz covariance (this is not Minkowski). Therefore, we employ (\ref{e3}) alone in the canonical quantization scheme, i.e. only one polarization $\hat \varepsilon \perp \v k_{\perp}$ interacts with the mirror and the other set of solutions for which $A_1 \neq 0$ becomes irrelevant. We write the commutators at equal times as

\bea
\left[ \hat \varepsilon \cdot \v A(\v x,t), g_{00}^{-1/2} \partial_t \hat \varepsilon \cdot \v A(\v x',t) \right] &=& i \delta^{3}(\v x - \v x') \nonumber \\
g^{00} \left[ \hat \varepsilon \cdot \v A(\v x,t), \hat \varepsilon \cdot \v A(\v x',t) \right] &=& 0.
\label{e7}
\eea
The mode decomposition of these operators rests on completeness relations, so first we define the 'free' modes as plane waves with appropriate metric normalization: $g^{1/4}e^{i\v k \cdot \v x }/(2\pi)^{3/2}$. Then we incorporte boundary conditions with the combination $\phi_q = g^{1/4} e^{-i\omega_q t + i\v q_{\perp} \cdot \v x}\sin\left[q(x-x_{\rm m}) \right]$, $\omega_q =\sqrt{q^2 + q_{\perp}^2}$ and $\v q_{\perp}=q_y \hat y + q_z \hat z$, all this for a single mirror. Using $\partial^0 = g^{00}\partial_0= (g_{00})^{-1}\partial_0 = (|g|)^{-1}\partial_0$ for the Rindler metric and $(d^3x \sqrt{|g|}) (\phi \partial^0 \psi) = (d^3x/ \sqrt{|g|} )(\phi \partial_0 \psi) $ for the volume element, the orthogonality of the following relativistic inner product is ensured

\bea
\<\phi_q, \phi_{q'} \> &=& i \int d^3 x \sqrt{|g|}(\phi_q^* \partial^0 \phi_{q'} - \phi_{q'} \partial^0 \phi_q^*) \nonumber \\
&=& -\frac{i}{(2\pi)^3} \int \frac{d^3 x} {\sqrt{|g|}} |g|^{1/4}|g|^{1/4} \times \nonumber \\ &\times& (i \omega_{q'}+i \omega_{q})e^{i(\v q_{\perp}-\v q'_{\perp}) \cdot \v x}  \times \nonumber \\
&\times& \sin\left[q (x-x_{\rm m}) \right] \sin\left[q' (x-x_{\rm m}) \right] \nonumber \\ &=& 2\omega_{q} \delta^{3}(\v q-\v q'). \nonumber \\
\<\phi_k, \phi^*_{\pm k'} \> &=& 0.
\label{e8}
\eea
For quantized frequencies $\omega_n$ of the accelerated eigenmodes, we have $\psi_{n,k_{\perp}} \equiv  e^{-i\omega_n t + i\v k_{\perp} \cdot \v x}\sqrt{k_{\perp}x} H^{(1)}_{i \omega/\alpha}(i k_{\perp} x) = e^{-i\omega_n t + i\v k_{\perp} \cdot \v x} \sqrt{\alpha x} \chi_n(x)$, where the specific form of $\omega_n(k_{\perp})$ will be discussed later. The inner product yields $\< \psi_n, \psi_m \> = 2 \omega_n(k_{\perp}) \delta_{n,m}\delta^{2}(\v k_{\perp}-\v k'_{\perp})$, and for negative energies or frequencies $\< \psi_n, \psi_m^* \>=0$. Finally, the Bogoliubov transformation from inertial operators $\hat A$ to non-inertial operators $\hat B$ involves both creation and annhilation of particles as 

\bea
\hat B_{n,k_{\perp}} = \int d^3 q &&\left[ \left(\frac{\< \psi_{n,k_{\perp}} , \phi_q \>^*}{2\sqrt{\omega_q \omega_n}} \right) \hat A_q \right. \nonumber \\ &+& \left. \left(\frac{\< \psi^*_{n,k_{\perp}} , \phi_q \>}{2\sqrt{\omega_q \omega_n}} \right) \hat A^{\dagger}_q  \right].
\label{e9}
\eea
The non-vanishing contribution of the transformed average number is controlled by the following overlap  

\bea
\< \phi_q^*, \psi_{n,k_{\perp}} \>
&=& e^{i(\omega_q-\omega_n)t} \left[\frac{\omega_n - \omega_q}{(2\pi)^{3/2}}\right] \times \nonumber \\ &\times& \int d^3 x e^{i(\v k_{\perp} - \v q_{\perp} )\cdot \v x} \sin\left[q(x-x_{\rm m})\right] \chi_n(x) \nonumber \\ &= & e^{i(\omega_q-\omega_n)t} \left[ \omega_n - \omega_q \right] \delta^{2}(\v k_{\perp} - \v q_{\perp}) F_{n,q}, \nonumber \\ F_{n,q}&\equiv& \frac{1}{\sqrt{2\pi}} \int_{x_{\rm m}}^{\infty} \sin\left[q(x-x_{\rm m})\right] \chi_n(x),
\label{e10}
\eea
where the complete set $\psi_{n,k_{\perp}} = e^{-i\omega_n t + i\v k_{\perp} \cdot \v x} g^{1/4} \chi_n(x)$ solves the 1D problem with anomalous potential and mirror boundary. One can see that $\chi_n$ is proportional to the Hankel function $H^{(1)}$; therefore, the remaining integral $F_{n,q}$ is the sine-Fourier transform of the corresponding Hankel function subject to (\ref{e5}). 
The expectation value of the number operator between inertial vacuum states $|0\>_A$ is not zero, which can be shown if we employ $ \hat A_{q'} \hat A_q^{\dagger} = \delta^{3}(\v q - \v q') +  \hat A_q^{\dagger} \hat A_{q'}$ in the expression for $\< N \>$:

\bea
&&\<0| B_{n,k_{\perp}}^{\dagger} B_{n,k_{\perp}} |0\>_{A} =\int d^3q \int d^3 q' \delta^{2}(\v q'_{\perp} - \v k_{\perp}) \times \nonumber \\ &\times& \delta^{2}(\v q_{\perp} - \v k_{\perp}) \delta^3(\v q- \v q') \frac{|F_{n,q}|^2 \left[\omega_n - \omega_q \right]^2 }{4 \omega_n \omega_q} \nonumber \\
&=&V_{k_{\perp}} \int dq \frac{|F_{n,q}|^2 \left[\omega_n - \omega_q \right]^2 }{4 \omega_n \omega_q} \bigg\rvert_{\v q_{\perp} = \v k_{\perp}},
\label{e11}
\eea
where $V_{k_{\perp}}$ is the volume of all available states $\v k_{\perp}$, thus avoiding the evaluation of $\delta(0)$ as costumary. A variety of techniques can be employed in the estimation of the integrals involved in this expression. The Fourier transform of the Hankel function is best estimated by means of the stationary phase approximation applied to its integral representation (\cite{grad}, pg. 915, sec. (8.421) formula 8):

\bea
H_{\nu}^{(1)}(xz) &=& \frac{z^{\nu}e^{-i\nu \pi/2}}{i \pi} \int_{0}^{\infty} dt e^{ix(t+z^2/t)/2}t^{-\nu-1} \nonumber \\
&=& \frac{1}{i\pi} \int_{-\infty}^{\infty} du \quad e^{-\nu u + i|z|x \sinh u}  \nonumber \\
&\approx&  \sqrt{\frac{8 i \pi}{\sqrt{|\nu|^2- (|z|x)^2}}} \cos \left[ |\nu| { \rm arccosh} \left(\frac{|\nu|}{|z|x} \right) \right. \nonumber \\ &-& \left. \sqrt{|\nu|^2-(|z|x)^2}- \pi/4 \right]\Theta\left(\left|\frac{\nu}{z} \right| -x \right).
\label{e12}
\eea

In passing, we note that this estimate resembles closely the WKB approximation of the 1D problem with anomalous $-1/x^2$ potential, but the singularity at the turning point $x=\sqrt{(\omega/\alpha)^2 + 1/4}/k_{\perp} \approx \omega/(k_{\perp}\alpha)$ can be 'cured' in the limit of small $\alpha$ and large $x \sim 1/\alpha$. We have $\sqrt{|\nu|-|z|^2 x^2}\approx (\sqrt{\omega^2 - k_{\perp}^2})/\alpha = k_{\parallel}/\alpha$. In the argument of the cosine we have $x \sqrt{(|\nu|/x)^2 - |z|^2} \approx x \sqrt{\omega^2 - k_{\perp}^2} = k_{\parallel} x$. Also in this limit, we have $ { \rm arccosh}(|\nu|/|z x|) \approx \log (2 |\nu|/|z x|)$. Regarding the normalization of bound solutions, we shall preserve the Hankel function's Dirac-delta convention satisfied by $\sqrt{z x}H_{\nu}^{(1)}(zx)$ instead of a Kronecker delta normalization of the square integrable function, because the limit $\alpha \rightarrow 0$, $x\rightarrow 1/\alpha \rightarrow \infty$ would make collapse all solutions $\psi_n$ to zero otherwise. Also, the argument of the cosine in (\ref{e12}) must be chosen such that $\psi_n(x_{\rm m})=0$. With these considerations, we have the following wave functions:

\bea
\psi_n &\approx&\sqrt{\frac{8\pi i k_{\perp} \alpha x}{k_{\parallel}}} \sin \left[ k_{\parallel} (x - x_{\rm m}) + \frac{\omega_n}{\alpha} \log \left( \frac{x}{x_{\rm m}} \right) \right] \times \nonumber \\ &\times& \Theta\left(\frac{\omega_n}{\alpha k_{\perp}} -x \right)
\label{e13}
\eea
while the implicit determination of the frequency function $\omega_n(k_{\perp})$ comes from the quantization condition (\ref{e5}) applied to (\ref{e12}):

\bea
&&\left(\frac{\omega}{\alpha} \right) {\rm arccosh\ }\left(\frac{\omega}{\alpha k_{\perp} x_{\rm m}} \right) - \sqrt{\left(\frac{\omega}{\alpha}\right)^2 - k_{\perp}^2 x_{\rm m}^2}= (n+1/4) \pi, \nonumber \\ &&\mbox{with} \quad  n=0,1,2,...
\label{e14}
\eea
A reasonable approximation for the frequency emerging from this relation is $\omega_n \approx \sqrt{(\alpha x_{\rm m}\pi (n+1/4) k_{\perp})^2 + k_{\perp}^2}$, thus quantizing $k_{\parallel} \approx k_{\perp}\alpha x_{\rm m}\pi (n+1/4)$, see fig. \ref{fig1}(d). With these approximations, the overlap $F_{n,q}$ can be put in terms of Gamma functions explicitly. For our purposes, it is better to estimate $F_{n,q}$ using a stationary phase approximation:

\bea
&&|F_{n,q}|^2 = \frac{k_{\perp} \omega_n }{k_{\parallel}|q-k_{\parallel}| \alpha}\times \nonumber \\ &\times& \Theta(q-k_{\parallel}-k_{\perp}) \Theta \left(k_{\parallel} +\frac{\omega}{\alpha x_{\rm m}} -q\right) \times \nonumber \\
&\times& \cos^2 \left[ \frac{\omega_n}{\alpha}\left\{1 + \log \left( \frac{\omega_n}{\alpha x_{\rm m} |q-k_{\parallel}|} \right)  \right\} -|q-k_{\parallel}| x_{\rm m} \right].\nonumber \\
\label{e15}
\eea
The computation of the limit $\alpha \rightarrow 0$ in this expression can be tricky, as $F_{n,q} \rightarrow \delta(q-k_{\parallel})$. This is important in the determination of (\ref{e11}) for low accelerations: The average photon number per mode satisfies $\< N \> \rightarrow 0$ as $\alpha \rightarrow 0$, which is verified by substituting $|F_{n,q}\times (\omega_n -\omega_q)|^2 \rightarrow |\delta(q-k_{\parallel}) (\omega_n -\omega_q)|^2=0 $. We also note that the location of the stationary point in the integral defining $F_{n,q}$ must satisfy

\bea
x^* =  \frac{\omega_n}{\alpha (q-k_{\parallel})}, \quad x_{\rm m}<x^*< \frac{\omega_n}{\alpha k_{\perp}}
\label{e15.1}
\eea
and it produces a natural infrared cutoff through the second inequality above. This is expressed in the first Heaviside Theta function in (\ref{e15}), since $q_{\rm min} = k_{\parallel} + k_{\perp}$ is the minimal parallel momentum of the photon. This also renders a nonsingular prefactor $1/(q-k_{\parallel})$ in (\ref{e15}). The second step function in (\ref{e15}) shows a natural ultraviolet cutoff $q_{\rm max}=k_{\parallel}+\omega_n /\alpha x_{\rm m}$, coming from the first inequality in (\ref{e15.1}). These cuts in the wavenumber also produce minimal and maximal photon frequencies $\omega_{\rm min} = \sqrt{q_{\rm min}^2 + k_{\perp}^2}$, $\omega_{\rm max} =\sqrt{q_{\rm max}^2 + k_{\perp}^2}$ respectively.

The integral over $q$ for the total average number per excited mode converges, and it is composed of an oscillatory (trigonometric) part and a monotonic function, as seen from the simple trick $\cos^2 \vartheta = (1+\cos 2\vartheta)/2$. We retain the strongest contribution from the non-oscillatory part of the integrand and find the following estimate:

\bea
\<N\> &\approx& \frac{V_{k_{\perp}}k_{\perp}}{8 k_{\parallel}} \left[ \omega_{\rm max}-\omega_{\rm min} + k_{\parallel} \log\left( \frac{q_{\rm max}+\omega_{\rm max}}{q_{\rm min}+\omega_{\rm min}}\right) \right. \nonumber \\  &-& \left. 2 \omega_n \log \left( \frac{k_{\perp}^2 + k_{\parallel}q_{\rm max} + \omega_n \omega_{\rm max}}{k_{\perp}^2 + k_{\parallel}q_{\rm min} + \omega_n \omega_{\rm min}} \right)\right].
\label{e16}
\eea
Some important conclusions can be drawn from our results. Firstly, the typical Boltzmann factor or Bose-Einstein distribution in the occupation numbers as functions of the energy are absent here; usually one finds such exponentials (and the corresponding effective temperatures in terms of $\alpha$) by computing the Bogoliubov coefficients through complex integration techniques {\it in the absence\ }of boundary conditions. Here, on the other hand, we have an important perturbation produced by the mirror. For very weak accelerations, we must consider mirrors located at $x_{\rm} \sim 1/\alpha$ and frontal photons such that $k_{\parallel} \gg k_{\perp}$. Then $\< N\> \sim a k_{\perp}/\alpha + b k_{\perp}^2 / \alpha k_{\parallel}$, with $a,b$ numerical constants. This means that the number decreases as $1/k_{\parallel}$, i.e. inversely proportional to photon's momentum. However, there is a remnant that vanishes linearly as $k_{\perp} \rightarrow 0$. No threshold mass or critical energy appeared in our calculations, as expected. As to the mirror's motion, it is clear now that the momentum imprinted on this object by the field is proportional to $\< \hat x \cdot \v E \times \v B \>$, which yields a superposition of all occupation numbers weighed by $k_{\parallel}$. Since the number per mode depends on $\alpha$ via $\omega_{n}$, the change of the mirror's linear momentum depends on the acceleration. Damping sets in, and there appears an effect similar to the Abraham-Lorentz force for classically accelerated charges, except that the mirror can be electrically neutral!
A concave mirror could act as a parachute, and help to manoeuvre a spaceship, in the event that it were ever possible to travel near the vicinity of a black hole. Plus the energy employed in the sudden acceleration of a photonic bottle can be converted into trapped radiation, a most attractive possibility. \vspace{6pt}

\begin{acknowledgments}
Financial support from VIEP project 100518931 BUAP-CA-289 is acknowledged. 
\end{acknowledgments}

\end{document}